\documentclass[twocolumn,showpacs,preprintnumbers,amsmath,amssymb,fleqn]{revtex4}

\usepackage{graphicx}
\usepackage{dcolumn}
\usepackage{bm}

\usepackage{color}
\usepackage{epsfig}

\newcommand{\gtapprox}{\raisebox{-0.5ex}{$\,\stackrel{>}{\scriptstyle\sim}\,$}}

\begin{document}



\title{Fermions in the pseudoparticle approach}

\author{Marc Wagner}

\affiliation{Institute~for~Theoretical~Physics~III, University~of~Erlangen-N{\"u}rnberg, Staudtstra{\ss}e~7, 91058~Erlangen, Germany}

\date{April~23, 2007}

\begin{abstract}
The pseudoparticle approach is a numerical technique to compute path integrals without discretizing spacetime. The basic idea is to integrate over those field configurations, which can be represented by a sum of a fixed number of localized building blocks (pseudoparticles). In a couple of previous papers we have successfully applied the pseudoparticle approach to pure SU(2) Yang-Mills theory. In this work we discuss how to incorporate fermionic fields in the pseudoparticle approach. To test our method, we compute the phase diagram of the 1+1-dimensional Gross-Neveu model in the large-$N$ limit.
\end{abstract}



\maketitle


\section{Introduction}

Recently, there have been a couple of papers proposing continuum models for SU(2) Yang-Mills theory with a small number of physically relevant degrees of freedom. There are ensembles of merons and regular gauge instantons \cite{Lenz:2003jp,Negele:2004hs}, there is the pseudoparticle approach \cite{Wagner:2005vs,Wagner:2006qn,Wagner:2006du} and there is a model with calorons with non trivial holonomy \cite{Gerhold:2006sk,Gerhold:2006kw}. The basic idea is to restrict the Yang-Mills path integral to those gauge field configurations, which can be represented by a linear superposition of a small number of localized building blocks, e.g.\ instantons, merons, akyrons or calorons. These models have been quite successful when dealing with problems related to confinement: the potential between two static charges is linear for large separations, there is a confinement deconfinement phase transition, and various quantities, e.g.\ the topological susceptibility or the critical temperature, are of the right order of magnitude compared to lattice results.

Until now, these models have been applied to SU(2) Yang-Mills theory only. In this paper we discuss how to include fermionic fields in the pseudoparticle approach.

The paper is organized as follows. In Sec.~\ref{SEC_001} we regularize the fermionic path integral by considering only those field configurations, which can be represented by a linear superposition of a fixed number of localized building blocks. We point out problems arising in a naive pseudoparticle regularization and propose a solution to these problems in form of a slightly different regularization scheme. We also discuss possible relations to finite mode regularization \cite{Andrianov:1982sn,Andrianov:1983fg,Andrianov:1983qj}. In Sec.~\ref{SEC_005} we test our pseudoparticle method by applying it to a simple interacting fermionic theory, the 1+1-dimensional Gross-Neveu model in the large $N$-limit \cite{Gross:1974jv}. With suitably chosen pseudoparticles and after a coupling constant renormalization our pseudoparticle results are in excellent agreement with analytical results, both for homogeneous chiral condensate \cite{Dashen:1974xz,Wolff:1985av} and for spatially inhomogeneous chiral condensate \cite{Thies:2003kk,Schnetz:2004vr}. In Sec.~\ref{SEC_008}, we give a summary and a brief outlook regarding the application of the pseudoparticle approach to QCD


\section{\label{SEC_001}Fermionic fields in the pseudoparticle approach}


\subsection{Basic principle}

In previous papers \cite{Wagner:2005vs,Wagner:2006qn,Wagner:2006du} we have discussed in detail, how to apply the pseudoparticle approach to bosonic fields, in particular to the SU(2) gauge field. In this section we propose a method to incorporate fermionic fields.
 
The starting point is action and partition function of any theory with quadratic fermion interaction:
\begin{eqnarray}
 & & \hspace{-0.44cm} S[\psi,\bar{\psi},\phi] \ \ = \ \ \int dx \, \Big(\bar{\psi} Q(\phi) \psi + \mathcal{L}(\phi)\Big) \\
 & & \hspace{-0.44cm} Z \ \ = \ \ \int D\psi \, D\bar{\psi} \, D\phi \, e^{-S[\psi,\bar{\psi},\phi]} ,
\end{eqnarray}
where $Q$ is the Dirac operator and $\phi$ denotes any type and number of bosonic fields, e.g.\ the chiral condensate in the Gross-Neveu model (cf.\ Sec.~\ref{SEC_006}) or the non-Abelian gauge field in QCD.

To keep close to the spirit of the pseudoparticle approach, we regularize the fermionic path integral by considering only those field configurations, which can be represented by a linear superposition a fixed number of localized building blocks:
\begin{eqnarray}
\psi(x) \ \ = \ \ \sum_j \underbrace{\eta_j G_j(x)}_{j\textrm{-th pseudoparticle}} ,
\end{eqnarray}
where $\eta_j$ are Grassmann valued spinors and $G_j$ are functions, which are localized in space as well as in time, i.e.\ pseudoparticles. The functional integration over the fermionic field configurations is defined via
\begin{eqnarray}
\int D\psi \, D\bar{\psi} \ldots \ \ = \ \ \int \Bigg(\prod_j d\eta_j \, d\bar{\eta}_j\Bigg) \ldots
\end{eqnarray}
Moreover, we consider a $d+1$-dimensional periodic spacetime region of temporal extension $L_0$ and spatial extension $(L_1)^d$. Note that fermionic fields have to fulfill antiperiodic boundary conditions in time direction (cf.\ e.g.\ \cite{Kapu89}). This implies antiperiodicity for the pseudoparticles $G_j$, i.e.\
\begin{eqnarray}
G_j(x_0+L_0,\mathbf{x}) \ \ = \ \ -G_j(x_0,\mathbf{x}) .
\end{eqnarray}
After integrating out the fermions one obtains
\begin{eqnarray}
\nonumber & & \hspace{-0.44cm} S_\textrm{effective}[\phi] \ \ = \\
\label{EQN_001} & & = \ \ \int d^{d+1}x \, \mathcal{L}(\phi) - \ln\Big(\det\Big(\langle G_j | Q | G_{j'} \rangle\Big)\Big) \\
 & & \hspace{-0.44cm} Z \ \ \propto \ \ \int D\phi \, e^{-S_\textrm{effective}[\phi]} ,
\end{eqnarray}
where $\langle G_j | Q | G_{j'} \rangle$ is a finite matrix.

Note that using eigenfunctions of the Dirac operator as ``pseudoparticles'' $G_j$ yields the well known finite mode regularization \cite{Andrianov:1982sn,Andrianov:1983fg,Andrianov:1983qj}.

If $\det(Q)$ is real and positive, $\det(Q) = \sqrt{\det(Q^\dagger Q)}$. This suggests another pseudoparticle regularization:
\begin{eqnarray}
\nonumber & & \hspace{-0.44cm} S_\textrm{effective}[\phi] \ \ = \\
\label{EQN_002} & & = \ \ \int d^{d+1}x \, \mathcal{L}(\phi) - \frac{1}{2} \ln\Big(\det\Big(\langle G_j | Q^\dagger Q | G_{j'} \rangle\Big)\Big) .
\end{eqnarray}
As we will point out in the following, this ``$Q^\dagger Q$-regular\-ization'' has significant advantages over the naive ``$Q$-reg\-ularization'' (\ref{EQN_001}).


\subsection{\label{SEC_002}The $Q$-regularization versus the $Q^\dagger Q$-regularization}

To keep the following arguments as simple as possible, we consider all pseudoparticles $G_j$ to be orthonormal, i.e.\ $\langle G_j | G_{j'} \rangle = \delta_{j j'}$. Note that assuming orthonormality is not a restriction. Given any set of linearly independent pseudoparticles $F_j$ one can easily define suitable linear combinations $G_j = M_{j k} F_k$, which are orthonormal. Up to an additive constant, which is irrelevant for the partition function, the result for the pseudoparticle regularized effective action is the same both for pseudoparticles $F_j$ and for pseudoparticles $G_j$:
\begin{eqnarray}
\nonumber & & \hspace{-0.44cm} \ln\Big(\det\Big(\langle G_j | Q | G_{j'} \rangle\Big)\Big) \ \ = \\
\nonumber & & = \ \ \ln\Big(\det\Big((M^\dagger)_{k j} \langle F_k | Q | F_{k'}\rangle M_{j' k'}\Big)\Big) \ \ = \\
 & & = \ \ \ln\Big(\det\Big(\langle F_j | Q | F_{j'} \rangle\Big)\Big) + \underbrace{\ln\Big(\det\Big(M^\dagger M\Big)\Big)}_{= \textrm{constant}} .
\end{eqnarray}


\subsubsection{\label{SEC_003}The Problem of the $Q$-regularization}

In the following we argue that the $Q$-regularization (\ref{EQN_001}) is not suited to produce physically meaningful results.

The problem of the $Q$-regularization is the following: applying the Dirac operator $Q$ to one of the pseudoparticles $G_{j'}$ in general yields a function, which is (partially) outside the pseudoparticle function space $\textrm{span}\{G_n\}$, i.e.\
\begin{eqnarray}
Q G_{j'}(x) \ \ = \ \ \sum_k a_{j' k} G_k(x) + h_{j'} H_{j'}(x)
\end{eqnarray}
with $H_{j'}$ normalized and $H_{j'} \perp \textrm{span}\{G_n\}$. If \\ $|\sum_k a_{j' k} G_k| \gg |h_{j'}|$, the situation is uncritical. However, as soon as $|h_{j'}|$ is of the same order of magnitude or even larger than $|\sum_k a_{j' k} G_k|$ problems arise: when computing the matrix elements $\langle G_j | Q | G_{j'} \rangle$ in (\ref{EQN_001}), a significant part of $Q G_{j'}$, $h_{j'} H_{j'}$, is simply ignored, just because $H_{j'}$ is perpendicular to the pseudoparticle function space $\textrm{span}\{G_n\}$.

If $Q$ is hermitian, the problem can be made even more transparent by diagonalizing the matrix $\langle G_j | Q | G_{j'} \rangle$:
\begin{eqnarray}
\nonumber & & \hspace{-0.44cm} (U^\dagger)_{j k} \langle G_k | Q | G_{k'}\rangle U_{k' j'} \ \ = \ \ \langle \tilde{G}_j | Q | \tilde{G}_{j'}\rangle \ \ = \\
 & & = \ \ \Big(\textrm{diag}(\mu_1 , \mu_2 , \ldots)\Big)_{j j'} ,
\end{eqnarray}
where $U$ is a unitary matrix and $\mu_j$ are the eigenvalues of $\langle G_j | Q | G_{j'}\rangle$. Because of $\det(U^\dagger U) = 1$,
\begin{eqnarray}
\det\Big(\langle G_j | Q | G_{j'} \rangle\Big) \ \ = \ \ \prod_j \mu_j .
\end{eqnarray}
Now the problem is obvious: according to
\begin{eqnarray}
\mu_j \ \ = \ \ \langle \tilde{G}_j | Q | \tilde{G}_j \rangle
\end{eqnarray}
$\tilde{G}_j$ contributes to the pseudoparticle regularized determinant of $Q$ as eigenmode with eigenvalue $\mu_j$. However, applying $Q$ to $\tilde{G}_j$ may yield a function, which is (partially) outside the pseudoparticle function space $\textrm{span}\{G_n\}$, i.e.\
\begin{eqnarray}
\label{EQN_003} Q \tilde{G}_j(x) \ \ = \ \ \mu_j \tilde{G}_j(x) + h_j H_j(x) .
\end{eqnarray}
If $|h_j| \gtapprox |\mu_j|$, $\tilde{G}_j$ is far from being an eigenfunction of $Q$, and $\mu_j$ is, of course, not related to any of the eigenvalues of $Q$.

The most extreme case is $\mu_j = 0$ and $h_j \neq 0$. $\tilde{G}_j$ is then an ``unphysical pseudoparticle zero mode'': although $\tilde{G}_j$ is not a zero mode of $Q$, and although $Q$ might not even have a zero mode, $\tilde{G}_j$ contributes to the pseudoparticle regularized determinant of $Q$ with eigenvalue \\ $\mu_j = 0$. Consequently, the effective action (\ref{EQN_001}) blows up to infinity. On the other hand, if we would use additional pseudoparticles with non vanishing overlap to $Q \tilde{G}_j$, $\tilde{G}_j$ would not enter the determinant as zero mode anymore.

Even if there are no unphysical pseudoparticle zero modes, there might still be unphysical low lying pseudoparticle modes, i.e.\ modes with $|h_j| \gtapprox |\mu_j|$, which also spoil numerical results.

The following example shows that such unphysical low lying pseudoparticle modes are quite common. The example in Sec.~\ref{SEC_004} and the $Q$-regularized pseudoparticle Gross-Neveu results from Sec.~\ref{SEC_007} demonstrate that these modes usually give rise to wrong and useless results.

\subsubsection*{A simple example}

We consider the antihermitian operator $\partial_x$, \\ $0 \leq x < L \in \{7 \, , \, 8\}$, antiperiodic boundary conditions.

The eigenfunctions of $\partial_x$ are plane waves and the corresponding eigenvalues are given by
\begin{eqnarray}
\label{EQN_004} \lambda_r \ \ = \ \ \frac{2 \pi i r}{L} \quad , \quad r \ \ = \ \ \ldots , -\frac{3}{2} , -\frac{1}{2} , +\frac{1}{2} , +\frac{3}{2} , \ldots
\end{eqnarray}

Alternatively, we compute the ``pseudoparticle eigenvalues'' $\mu_j$ and $|h_j|$ defined in (\ref{EQN_003}), where we use overlapping ``hat functions'' as pseudoparticles. To be more precise, we apply antiperiodic B-spline basis functions of degree $2$, i.e.\
\begin{eqnarray}
G_j(x) \ \ = \ \ B_{j,\textrm{antiperiodic}}^{(2)}(x) ,
\end{eqnarray}
$j = 0,\ldots,L-1$ (cf.\ Appendix~\ref{SEC_009}).

In Table~\ref{TAB_001} we compare the ``true eigenvalues'' $\lambda_r$ with the pseudoparticle eigenvalues $\mu_j$. We also show $|h_j|$, the norm of the overlap of $\partial_x \tilde{G}_j$ to the function space perpendicular to $\textrm{span}\{G_n\}$. As we have discussed, the pseudoparticle eigenvalues are quite similar to certain true eigenvalues as long as $|\mu_j| \gg |h_j|$. However, as soon as they are of the same order of magnitude, unphysical low lying pseudoparticle modes or, in the case of odd $L$, unphysical pseudoparticle zero modes appear.

\begin{table}[h!]
\begin{center}
\begin{tabular}{|c|c|c|c||c|c|c|c|}
\hline
\multicolumn{4}{|c||}{$L=7$} & \multicolumn{4}{c|}{$L=8$} \\
\hline
$r$ & $\lambda_r$ & $\mu_j$ & $|h_j|$ & $r$ & $\lambda_r$ & $\mu_j$ & $|h_j|$ \\
\hline
1/2 & $0.449 \, i$ & $\pm 0.449 \, i$ & 0.004 & 1/2 & $0.393 \, i$ & $\pm 0.393 \, i$ & 0.002 \\
3/2 & $1.346 \, i$ & $\pm 1.344 \, i$ & 0.135 & 3/2 & $1.178 \, i$ & $\pm 1.177 \, i$ & 0.083 \\
5/2 & $2.244 \, i$ & $\pm 2.065 \, i$ & 1.063 & 5/2 & $1.963 \, i$ & $\pm 1.909 \, i$ & 0.599 \\
7/2 & $3.142 \, i$ & $0.000$      & 3.162 & 7/2 & $2.749 \, i$ & $\pm 1.610 \, i$ & 2.441 \\
\hline
\end{tabular}
\caption{\label{TAB_001}True eigenvalues $\lambda_r$, pseudoparticle eigenvalues $\mu_j$ and $|h_j|$.}
\end{center}
\end{table}

Note that such unphysical pseudoparticle zero modes are not specific for B-spline pseudoparticles. For odd $L$ and any choice of localized, real-valued and uniformly distributed pseudoparticles $G_j$, i.e.\
\begin{eqnarray}
G_j(x) \ \ = \ \ F(x-j) - F(x-j+L) ,
\end{eqnarray}
$j = 0,\ldots,L-1$, with $F^\ast = F$ and $F = 0$ for $x \leq 0$ and $x \geq L$, one can easily show that
\begin{eqnarray}
\tilde{G}_\textrm{zero mode}(x) \ \ = \ \ \sum_j (-1)^j G_j(x)
\end{eqnarray}
is perpendicular to $\textrm{span}\{G_n\}$. Therefore, $\tilde{G}_\textrm{zero mode}$ is an unphysical pseudoparticle zero mode of $\partial_x$.


\subsubsection{The advantage of the $Q^\dagger Q$-regularization}

The matrix elements $\langle G_j | Q^\dagger Q | G_{j'} \rangle$ in the $Q^\dagger Q$-regu\-larized effective action (\ref{EQN_002}) do not suffer from the problem discussed in the previous section. The reason is the following: both the left hand sides $\langle G_j | Q^\dagger$ and the right hand sides $Q | G_{j'} \rangle$ might be outside the pseudoparticle function space $\textrm{span}\{G_n\}$, but they form the same function space, $\textrm{span}\{Q G_n\}$, in which their overlap is computed.

For example, it is easy to show that any pseudoparticle zero mode of $Q^\dagger Q$ is necessarily a ``true zero'' mode of $Q^\dagger Q$: if $G_\textrm{zero mode}$ is a pseudoparticle zero mode of $Q^\dagger Q$, then $\langle G_\textrm{zero mode} | Q^\dagger Q | G_\textrm{zero mode} \rangle = 0$; this implies $Q | G_\textrm{zero mode} \rangle = 0$ and $Q^\dagger Q | G_\textrm{zero mode} \rangle = 0$.


\subsubsection{Another way to motivate the $Q^\dagger Q$-regularization}

It is easy to show that using complete but possibly different orthonormal sets of functions $\{F_n\}$ and $\{G_n\}$ on the left hand side and on the right hand side of any operator $Q$ to calculate its determinant yields the determinant with exception of a phase factor, i.e.\
\begin{eqnarray}
\label{EQN_005} \Big|\det(Q)\Big| \ \ = \ \ \left|\det\Big(\langle F_j | Q | G_{j'} \rangle\Big)\right| .
\end{eqnarray}

As we have already stressed, the problem of the $Q$-regularization (\ref{EQN_001}) is that $Q G_{j'}$ is (partially) perpendicular to $\textrm{span}\{G_n\}$. On the other hand, when considering operators, where $\det(Q)$ is real and positive, we are only interested in the absolute value of the determinant. Therefore, according to (\ref{EQN_005}) we propose to consider matrix elements with different ``pseudoparticles'' on the left hand side and on the right hand side:
\begin{itemize}
\item Right hand side: any choice of pseudoparticles $G_{j'}$.

\item Left hand side: ``pseudoparticles'' $F_j$ with \\ $\textrm{span}\{F_n\} = \textrm{span}\{Q G_n\}$.
\end{itemize}
This assures that $Q$ applied to any pseudoparticle $G_{j'}$ yields an element of the left hand side function space. In other words, the best choice for the left hand side is to use basis functions $F_j$ with $\textrm{span}\{F_n\} = \textrm{span}\{Q G_n\}$, because other perpendicular/partially perpendicular basis functions have no/not enough overlap to $Q G_{j'}$, and this in turn leads to unphysical pseudoparticle zero modes/low lying modes.

One can show that
\begin{eqnarray}
\Big|\det\Big(\langle F_j | Q | G_{j'} \rangle\Big)\Big| \ \ = \ \ \sqrt{\det\Big(\langle G_j | Q^\dagger Q | G_{j'} \rangle\Big)} ,
\end{eqnarray}
i.e.\ this line of reasoning also leads to the $Q^\dagger Q$-regular\-ization (\ref{EQN_002}).


\subsection{\label{SEC_004}The $Q^\dagger Q$-regularization and its relation to finite mode regularization}

In the following we point out that there are certain relations between the $Q^\dagger Q$-regularization of the pseudoparticle approach and finite mode regularization \cite{Andrianov:1982sn,Andrianov:1983fg,Andrianov:1983qj}.

Let $\psi_n$ be orthonormalized eigenfunctions and $\lambda_n$ the corresponding eigenvalues of $Q^\dagger Q$, i.e.\
\begin{eqnarray}
\label{EQN_006} Q^\dagger Q \psi_n \ \ = \ \ \lambda_n \psi_n .
\end{eqnarray}
Of course, $\lambda_n$ is real and $\lambda_n \geq 0$, because of
\begin{eqnarray}
\label{EQN_007} \lambda_n \ \ = \ \ \langle \psi_n | Q^\dagger Q | \psi_n \rangle \ \ = \ \ \Big| Q | \psi_n \rangle \Big|^2 \ \ \geq \ \ 0 .
\end{eqnarray}
In the following we consider the eigenvalues ordered according to their absolute value, i.e.\ $|\lambda_0| < |\lambda_1| < \ldots$ Moreover, we differentiate between ``low lying eigenvalues'' $\lambda_0,\ldots,\lambda_{M-1}$ and ``large eigenvalues'' $\lambda_M,\ldots$:
\begin{itemize}
\item \textbf{Large eigenvalues} \\ In applications of the pseudoparticle approach, bosonic fields $\phi$ have typical maximum values (cf.\ e.g.\ \cite{Wagner:2006qn}). The reason is both the pseudoparticle regularization and the exponential damping factor $e^{-S}$ in the partition function. The same is true for the ``potential'' $V(\phi)$ in the Dirac operator \\ $Q = \gamma_\mu \partial_\mu + V(\phi)$. Therefore, plane waves $\eta e^{i k x}$ with large wave numbers $k$ are approximate eigenfunctions of $Q^\dagger Q$ (``the derivative dominates the potential''). Both the eigenfunctions $\psi_n \approx \eta e^{i k x}$ and the corresponding eigenvalues $\lambda_n \approx i k$, \\ $n = M,\ldots$, are nearly independent of the bosonic fields $\phi$.

\item \textbf{Low lying eigenvalues} \\ Low lying eigenvalues $\lambda_n$ and the corresponding eigenfunctions $\psi_n$, $n = 0,\ldots,M-1$, may exhibit a strong $\phi$-dependence. Because the bosonic fields $\phi$ have typical maximum values, there is only a finite number of low lying eigenvalues.
\end{itemize}

Finite mode regularization of $\det(Q^\dagger Q)$ amounts to considering only a finite number of low lying eigenvalues, i.e.\
\begin{eqnarray}
\label{EQN_008} \det(Q^\dagger Q) \ \ \textrm{``}=\textrm{''} \ \ \prod_{j=0}^{N-1} \lambda_j .
\end{eqnarray}
Of course, this equation is not an equality in the usual sense. The hope is rather that physical observables can be computed correctly via the partition function, when using (\ref{EQN_008}) in the effective action, and when performing a suitable renormalization,

To exhibit the relation between the $Q^\dagger Q$-regularization of the pseudoparticle approach and finite mode regularization, it is convenient to diagonalize the matrix \\ $\langle G_j | Q^\dagger Q | G_{j'} \rangle$:
\begin{eqnarray}
\nonumber & & \hspace{-0.44cm} (U^\dagger)_{j k} \langle G_k | Q^\dagger Q | G_{k'}\rangle U_{k' j'} \ \ = \ \ \langle \tilde{G}_j | Q^\dagger Q | \tilde{G}_{j'}\rangle \ \ = \\
\label{EQN_009} & & = \ \ \Big(\textrm{diag}(\mu_1 \, , \, \mu_2 \, , \, \ldots)\Big)_{j j'} ,
\end{eqnarray}
where $U$ is a unitary matrix and $\mu_j$ are the eigenvalues of $\langle G_j | Q^\dagger Q | G_{j'} \rangle$. Because of $\det(U^\dagger U) = 1$,
\begin{eqnarray}
\label{EQN_010} \det\Big(\langle G_j | Q^\dagger Q | G_{j'} \rangle\Big) \ \ = \ \ \prod_j \mu_j .
\end{eqnarray}

In the following we assume that the pseudoparticles $G_j$ have been chosen such that the low lying eigenfunctions $\psi_n$, $n = 0,\ldots,M-1$, can be approximated. Then
\begin{eqnarray}
\tilde{G}_j \ \ \approx \ \ \psi_j
\end{eqnarray}
and
\begin{eqnarray}
\mu_j \ \ = \ \ \langle \tilde{G}_j | Q^\dagger Q | \tilde{G}_j \rangle \ \ \approx \ \ \lambda_j ,
\end{eqnarray}
$j = 0,\ldots,M-1$. That is, the low lying, $\phi$-dependent pseudoparticle eigenfunctions $\tilde{G}_j$ and the pseudoparticle eigenvalues $\mu_j$ are nearly identical to the true eigenfunctions $\psi_j$ and eigenvalues $\lambda_j$. Therefore, the contribution of these pseudoparticle eigenmodes to the pseudoparticle regularized determinant of $Q^\dagger Q$ is identical to the contribution of the corresponding true eigenmodes in finite mode regularization (cf.\ (\ref{EQN_008}) and (\ref{EQN_010})).

All other pseudoparticle eigenfunctions are essentially linear combinations of the remaining true eigenfunctions:
\begin{eqnarray}
\label{EQN_011} \tilde{G}_j \ \ \approx \ \ \sum_{n=M}^\infty c_{j n} \psi_n \quad , \quad \sum_{n=M}^\infty \Big|c_{j n}\Big|^2 \ \ = \ \ 1 ,
\end{eqnarray}
$j = M,\ldots$ (there is no contribution from $\psi_0,\ldots,\psi_{M-1}$, because the pseudoparticle eigenmodes are orthonormal due to the hermiticity of $\langle G_j | Q^\dagger Q | G_{j'} \rangle$). The corresponding pseudoparticle eigenvalues $\mu_j$ are given by
\begin{eqnarray}
\label{EQN_012} \mu_j \ \ = \ \ \langle \tilde{G}_j | Q^\dagger Q | \tilde{G}_j \rangle \ \ \approx \ \ \sum_{n=M}^\infty \Big|c_{j n}\Big|^2 \lambda_n \ \ \geq \ \ \lambda_M ,
\end{eqnarray}
$j = M,\ldots$, where (\ref{EQN_007}) and (\ref{EQN_011}) have been used. There are two important points:
\begin{itemize}
\item $\tilde{G}_j$ and $\mu_j$, $j = M,\ldots$, are approximately independent of $\phi$. That is, although the difference between these pseudoparticle eigenvalues $\mu_j$ and the true eigenvalues $\lambda_j$, $j = M,\ldots$, might be large, the pseudoparticle effective action merely differs by an additive constant, when compared to finite mode regularization. Such an additive constant is, of course, irrelevant for the partition function.

\item $\mu_j \geq \lambda_M$, $j = M,\ldots$, i.e.\ $\mu_j$ is large. Therefore, possibly present weak $\phi$-dependencies of $c_{j n}$ or $\lambda_j$ or weak contributions from low lying eigenmodes $\psi_n$, $n = 0,\ldots,M-1$, do not have a strong impact on $\mu_j$. Assuming that there are not too many of these large pseudoparticle eigenmodes, their $\phi$-de\-pendence is essentially negligible in the pseudoparticle regularized determinant of $Q^\dagger Q$.
\end{itemize}

It is also instructive to discuss the shortcoming of the $Q$-regularization in this context. For hermitian $Q$ the above arguments and equations are quite similar: just replace $Q^\dagger Q$ by $Q$ in (\ref{EQN_006}) to (\ref{EQN_011}) with exception of (\ref{EQN_007}), which is of course not valid. (\ref{EQN_012}) must be replaced by
\begin{eqnarray}
\mu_j \ \ = \ \ \langle \tilde{G}_j | Q | \tilde{G}_j \rangle \ \ \approx \ \ \sum_{n=M}^\infty \Big|c_{j n}\Big|^2 \lambda_n .
\end{eqnarray}
The big difference compared to the $Q^\dagger Q$-regularization is that the true eigenvalues $\lambda_j$ are, in general, not positive. Therefore, the corresponding pseudoparticle eigenvalues $\mu_j$, $j = M,\ldots$, are not necessarily large: due to intricate cancellations between positive and negative $\lambda_j$, there may very well be small pseudoparticle eigenvalues $\mu_j$, which are strongly affected by small changes of $c_{j n}$ or $\lambda_j$, because of a change of $\phi$.

\subsubsection*{Another simple example}

In this example we compare the $Q^\dagger Q$-regularization, the $Q$-regularization and finite mode regularization. To this end, we compute
\begin{eqnarray}
\ln\Big(\det\Big(\partial_x + \sigma\Big)\Big) ,
\end{eqnarray}
$\sigma$ real, $0 \leq x < L = 7$, antiperiodic boundary conditions.

\begin{figure}[t!]
\begin{center}
\includegraphics{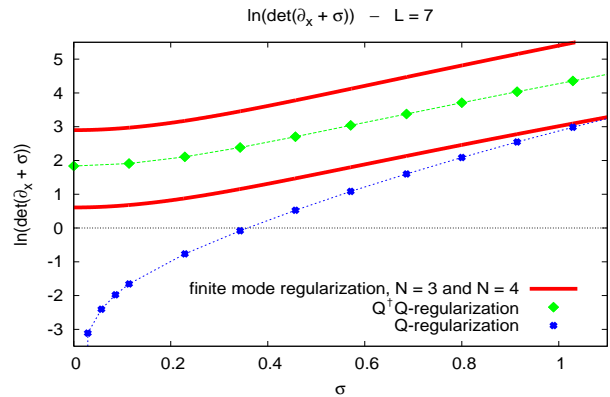}
\caption{\label{FIGURE_001}$L = 7$. $\ln(\det(\partial_x + \sigma))$ as function of $\sigma$ computed via finite mode regularization, $Q^\dagger Q$-regularization and $Q$-reg\-ularization.}
\end{center}
\end{figure}

\begin{itemize}
\item \textbf{Finite mode regularization} \\ The eigenfunctions of $\partial_x + \sigma$ are plane waves and the corresponding eigenvalues are given by
\begin{eqnarray}
\nonumber & & \hspace{0.5cm} \lambda_r \ \ = \ \ \frac{2 \pi i r}{L} + \sigma \quad , \\
 & & \hspace{1.65cm} r \ \ = \ \ \ldots , -\frac{3}{2} , -\frac{1}{2} , +\frac{1}{2} , +\frac{3}{2} , \ldots
\end{eqnarray}
According to (\ref{EQN_008}) the result is
\begin{eqnarray}
\nonumber & & \hspace{0.5cm} \ln\Big(\det\Big(\partial_x + \sigma\Big)\Big) \ \ = \\
 & & \hspace{1.65cm} = \ \ \sum_{n=0}^{N-1} \ln\left(\left(\frac{2 \pi (n+1/2)}{L}\right)^2 + \sigma^2\right) ,
\end{eqnarray}
when using the $2 N$ ``lowest lying eigenmodes''. For $N = 3$ and $N = 4$ it is shown as a function of $\sigma$ in Fig.~\ref{FIGURE_001}.

\item $Q^\dagger Q$\textbf{-regularization and }$Q$\textbf{-regularization} \\ We use orthonormalized B-spline pseudoparticles of degree $2$ as pseudoparticles $G_j$. The $Q^\dagger Q$-result,
\begin{eqnarray}
 & & \hspace{0.5cm} \frac{1}{2} \ln\Big(\det\Big(\langle G_j | \Big(\partial_x + \sigma\Big)^\dagger \Big(\partial_x + \sigma\Big) | G_{j'} \rangle\Big)\Big) ,
\end{eqnarray}
and the $Q$-result,
\begin{eqnarray}
 & & \hspace{0.5cm} \ln\Big(\det\Big(\langle G_j | \Big(\partial_x + \sigma\Big) | G_{j'} \rangle\Big)\Big) ,
\end{eqnarray}
are shown as functions of $\sigma$ in Fig.~\ref{FIGURE_001}.
\end{itemize}

The results obtained with $Q^\dagger Q$-regularization and with finite mode regularization are nearly identical with exception of a shift along the vertical axis. This shift is due to the different number of degrees of freedom: $7$ for the pseudoparticle regularization, $6$ and $8$ for finite mode regularization. The $Q$-regularized result, on the other hand, has a completely different shape, especially for small values of $\sigma$. As explained above, the reason are unphysical low lying pseudoparticle modes. At $\sigma = 0$ an unphysical pseudoparticle zero mode causes a singularity.

\subsubsection*{Computing approximate eigenfunctions via the pseudoparticle approach}

The pseudoparticle approach can also be used to compute approximate eigenfunctions of $Q^\dagger Q$ (if $Q$ is hermitian, one can also use $Q$ instead of $Q^\dagger Q$):
\begin{itemize}
\item Let $G_j$ be a set of orthonormalized pseudoparticles.

\item Compute the matrix $\langle G_j | Q^\dagger Q | G_{j'} \rangle$ and diagonalize it according to (\ref{EQN_009}).

\item Compute $|h_j|$, defined by
\begin{eqnarray}
 & & \hspace{0.5cm} Q^\dagger Q \tilde{G}_j(x) \ \ = \ \ \mu_j \tilde{G}_j(x) + h_j H_j(x)
\end{eqnarray}
with $H_j$ normalized and $H_j \perp \textrm{span}\{G_n\}$.

\item If $|\mu_j| \gg |h_j|$, the pseudoparticle eigenfunction $\tilde{G}_j$ is close to a true eigenfunction of $Q^\dagger Q$ and the corresponding pseudoparticle eigenvalue $\mu_j$ is close to a true eigenvalue.

\item If $|\mu_j| \approx |h_j|$ or if $|\mu_j| < |h_j|$, the pseudoparticle eigenfunction $\tilde{G}_j$ is not close to a true eigenfunction of $Q^\dagger Q$.
\end{itemize}

\subsubsection*{The simple example continued}

We continue the example of Sec.~\ref{SEC_003} by computing the pseudoparticle eigenfunctions $\tilde{G}_j$ of the antihermitian operator $\partial_x$ for $L = 7$.

\begin{figure}[h!]
\begin{center}
\includegraphics{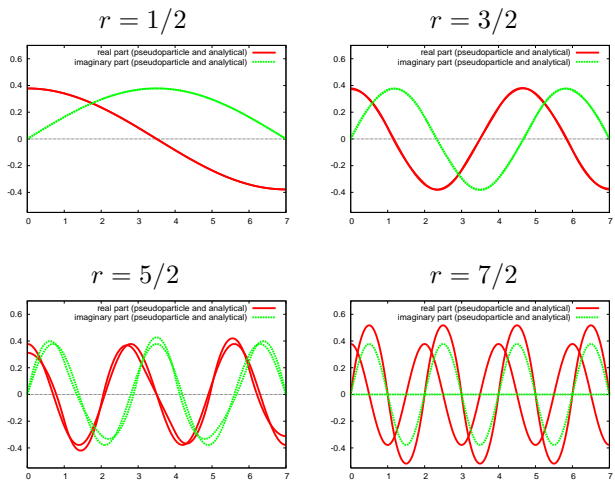}
\caption{\label{FIGURE_002}$L = 7$. Real parts and imaginary parts of the pseudoparticle eigenfunctions and the true eigenfunctions of $\partial_x$ as functions of $x$.}
\end{center}
\end{figure}

In Fig.~\ref{FIGURE_002} we compare both the real parts and the imaginary parts of the pseudoparticle eigenfunctions and the true eigenfunctions $(1 / \sqrt{L}) e^{\lambda_r x}$ with $\lambda_r$ given by (\ref{EQN_004}). As long as $|\mu_j| \gg |h_j|$ ($r = 1/2$ and $r = 3/2$; cf.\ Table~\ref{TAB_001}) pseudoparticle results and analytical results are essentially identical. However, if $|\mu_j| < |h_j|$ ($r = 7/2$), $\tilde{G}_j$ is not close to a true eigenfunction of $\partial_x$ anymore.

\section{\label{SEC_005}The 1+1-dimensional Gross-Neveu model in the pseudoparticle approach}

In the following, we test the pseudoparticle approach by considering a simple interacting fermionic theory, the 1+1-dimensional Gross-Neveu model in the large-$N$ limit. We compute the phase diagram both for homogeneous and for spatially inhomogeneous chiral condensate.


\subsection{\label{SEC_006}The 1+1-dimensional Gross-Neveu model in Euclidean spacetime}

The Gross-Neveu model \cite{Gross:1974jv} is a four fermion interacting theory with $N$ degenerate flavors. Action and partition function are given by
\begin{eqnarray}
\nonumber & & \hspace{-0.44cm} S \ \ = \ \ \int d^2x \, \Bigg(\sum_{n=1}^N \bar{\psi}^{(n)} \Big(\gamma_0 (\partial_0 + \mu) + \gamma_1 \partial_1\Big) \psi^{(n)} - \\
 & & \hspace{0.62cm} \frac{g^2}{2} \left(\sum_{n=1}^N \bar{\psi}^{(n)} \psi^{(n)}\right)^2\Bigg) \\
 & & \hspace{-0.44cm} Z \ \ = \ \ \int \left(\prod_{n=1}^N D\psi^{(n)} \, D\bar{\psi}^{(n)}\right) e^{-S} ,
\end{eqnarray}
where $N$ is the number of flavors, $\mu$ is the chemical potential and $g$ is the coupling constant. The Dirac matrices fulfill $\{\gamma_\mu , \gamma_\nu\} = 2 \delta_{\mu \nu}$, e.g.\ $\gamma_0 = \sigma^1$ and $\gamma_1 = \sigma^3$.

To get rid of the four fermion interaction, one usually introduces a real scalar field $\sigma$:
\begin{eqnarray}
\nonumber & & \hspace{-0.44cm} S' \ \ = \ \ \int d^2x \, \Bigg(\frac{1}{2 g^2} \sigma^2 +
\end{eqnarray}
%
%
\begin{eqnarray}
 & & \hspace{0.62cm} \sum_{n=1}^N \bar{\psi}^{(n)} \underbrace{\Big(\gamma_0 (\partial_0 + \mu) + \gamma_1 \partial_1 + \sigma\Big)}_{= Q \quad \textrm{(Dirac operator)}} \psi^{(n)}\Bigg) \\
 & & \hspace{-0.44cm} Z \ \ \propto \ \ \int \left(\prod_{n=1}^N D\psi^{(n)} \, D\bar{\psi}^{(n)}\right) \int D\sigma \, e^{-S'} .
\end{eqnarray}
Integrating out the fermions yields
\begin{eqnarray}
 & & \hspace{-0.44cm} S_\textrm{effective} \ \ = \ \ N \left(\frac{1}{2 \lambda} \int d^2x \, \sigma^2 - \ln\Big(\det(Q)\Big)\right) \\
\label{EQN_013} & & \hspace{-0.44cm} Z \ \ \propto \ \ \int D\sigma \, e^{-S_\textrm{effective}} ,
\end{eqnarray}
where $\lambda = N g^2$.

From now on we consider the large-$N$ limit, i.e.\ \\ $N \rightarrow \infty$ and $\lambda = \textrm{constant}$. Note that due to \\ $S_\textrm{effective} \propto N$, only a single field configuration contributes to the partition function (\ref{EQN_013}). It can be determined by minimizing $S_\textrm{effective}$ with respect to $\sigma$.

Moreover, one can show that in the large-$N$ limit
\begin{eqnarray}
\sigma \ \ = \ \ -g^2 \sum_{n=1}^N \bar{\psi}^{(n)} \psi^{(n)} ,
\end{eqnarray}
i.e.\ the scalar field $\sigma$ is proportional to the chiral condensate.


\subsection{B-spline pseudoparticles}

\begin{figure}[h!]
\begin{center}
\includegraphics{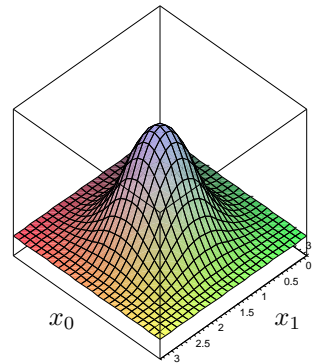}
\caption{\label{FIGURE_003}B-spline pseudoparticle $G_{0,0}$ as a function of $(x_0, x_1)$.}
\end{center}
\end{figure}

For the following computations we use a large number of overlapping ``hat functions'' as pseudoparticles: we apply products of antiperiodic and periodic B-spline basis functions of degree $2$, i.e.\
\begin{eqnarray}
G_{j_0,j_1}(x_0,x_1) \ \ = \ \ B_{j_0,\textrm{antiperiodic}}^{(2)}(x_0) B_{j_1,\textrm{periodic}}^{(2)}(x_1)
\end{eqnarray}
$j_0 = 0,\ldots,L_0-1$, $j_1 = 0,\ldots,L_1-1$ (cf.\ Appendix~\ref{SEC_009}), where $L_0 \times L_1$ is the extension of the periodic spacetime region ($L_0$ and $L_1$ are chosen to be integers). Fig.~\ref{FIGURE_003} shows the ``B-spline pseudoparticle'' $G_{0 0}$.

\subsubsection*{Why using B-spline pseudoparticles?}

\begin{itemize}
\item The intention of this section is to test, whether the pseudoparticle approach as presented in the previous section is suited to deal with fermionic fields. Therefore, we need pseudoparticles, which form a ``sensible set of field configurations'', i.e.\ pseudoparticles, which are able to approximate any not too heavily oscillating field configuration. For B-spline pseudoparticles, i.e.\ for a piecewise polynomial basis, this is certainly the case. Note that it is not our intention to determine a small number of physically important fermionic field configurations; this will be part of an upcoming paper, where we apply the pseudoparticle approach to QCD.

\item Since B-spline pseudoparticles are piecewise polynomial functions, the matrix elements \\ $\langle G_{j_0,j_1} | Q^\dagger Q | G_{j'_0,j'_1} \rangle$ can be calculated analytically.

\item B-spline pseudoparticles of degree $2$ with uniform knot vectors $t_j = j$ are localized within a spacetime region of extension $3 \times 3$. Therefore, when using a large number of pseudoparticles the matrix $\langle G_{j_0,j_1} | Q^\dagger Q | G_{j'_0,j'_1} \rangle$ is sparse, which is beneficial from a numerical point of view.
\end{itemize}


\subsection{\label{SEC_007}Homogeneous chiral condensate}

In this section we consider a homogeneous chiral condensate, i.e.\ $\sigma = \textrm{constant}$.

The $Q^\dagger Q$-regularized pseudoparticle effective action is given by
\begin{eqnarray}
\nonumber & & \hspace{-0.44cm} \frac{S_\textrm{effective}}{N} \ \ = \ \ \frac{1}{2 \lambda} \int d^2x \, \sigma^2 - \\
\label{EQN_014} & & \hspace{0.62cm} \frac{1}{2} \ln\Big(\det\Big(\langle G_{j_0,j_1} | Q^\dagger Q | G_{j'_0,j'_1} \rangle\Big)\Big) .
\end{eqnarray}
To determine the chiral condensate $\sigma$ for given temperature $T = 1 / L_0$ and chemical potential $\mu$, one has to minimize this expression with respect to $\sigma$.

\begin{figure}[t!]
\begin{center}
\includegraphics{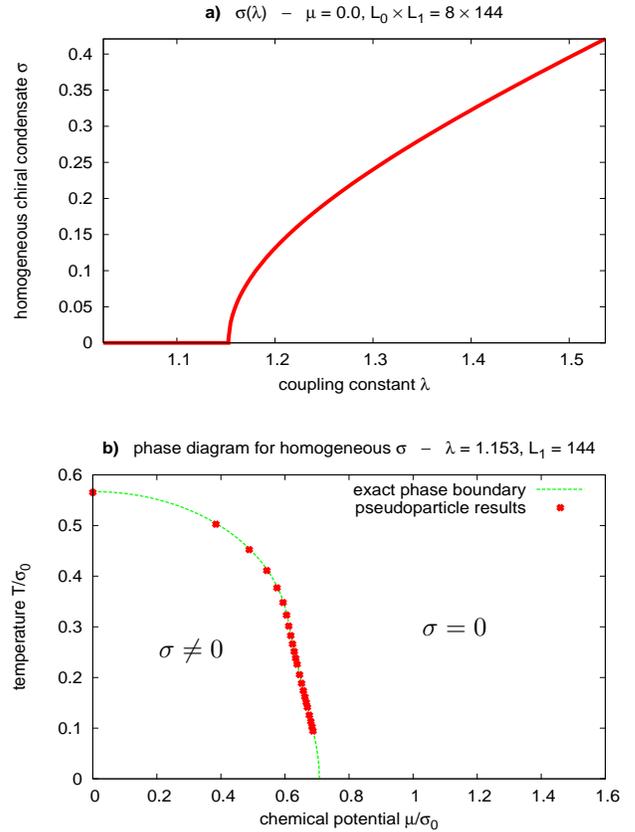}
\caption{\label{FIGURE_004}$Q^\dagger Q$-regularization, $L_1 = 144$.
\textbf{a)}~$\mu = 0.0$, $L_0 = 8$. $\sigma$ as a function of $\lambda$.
\textbf{b)}~$\lambda = 1.153$. The pseudoparticle phase diagram for homogeneous chiral condensate together with the exact phase boundary.
}
\end{center}
\end{figure}

Of course, numerical results strongly depend on the number of pseudoparticles applied. To extract physically meaningful results, a coupling constant renormalization is necessary, i.e.\ we have to choose $\lambda$ in accordance with the number of degrees of freedom. A possible way of doing that is to perform finite temperature computations with temporal extension $L_0 = 8$ much smaller than spatial extension $L_1 = 144$ at $\mu = 0.0$ for different values of $\lambda$. The resulting $\sigma$ as a function of $\lambda$ is shown in Fig.~\ref{FIGURE_004}a. For all further computations we fix the coupling constant $\lambda$ at that value, where the chiral condensate $\sigma$ just vanishes: $\lambda = 1.153$. The scale is now set, i.e.\ $1/L_0 = 1/8$ corresponds to the critical temperature of chiral symmetry breaking at $\mu = 0.0$.

We can now compute the chiral condensate $\sigma$ at ``arbitrary'' temperature $T = 1 / L_0$ ($L_0$ integer) and chemical potential $\mu$, to determine the phase diagram of the Gross-Neveu model. As usual, we express $T$ and $\mu$ in terms of $\sigma_0$, which is the $T = 0.0$ and $\mu = 0.0$ value of the chiral condensate $\sigma$ (to determine $\sigma_0$ we have performed a ``zero temperature computation'' with large temporal extension $L_0 = 48$). The resulting pseudoparticle phase diagram is in excellent agreement with analytical results \cite{Dashen:1974xz,Wolff:1985av} (cf.\ Fig.~\ref{FIGURE_004}b).

Another possibility to adjust the temperature $T / \sigma_0$ is to change the coupling constant $\lambda$, while the extension of the spacetime region $L_0 \times L_1$ is kept constant. This is similar to what is usually done in lattice calculations. The scale is then set via $\sigma_0$, which is $\lambda$-dependent. Using this method the resulting pseudoparticle phase diagram is also in excellent agreement with analytical results.

In Sec.~\ref{SEC_002} we have pointed out that the $Q$-regulariza\-tion is not suited to produce physically meaningful results. To demonstrate that this is indeed the case, we perform similar computations with the $Q$-regularized version of the effective action,
\begin{eqnarray}
\nonumber & & \hspace{-0.44cm} \frac{S_\textrm{effective}}{N} \ \ =
\end{eqnarray}
%
%
\begin{eqnarray}
 & & = \ \ \frac{1}{2 \lambda} \int d^2x \, \sigma^2 - \ln\Big(\det\Big(\langle G_{j_0,j_1} | Q | G_{j'_0,j'_1} \rangle\Big)\Big) .
\end{eqnarray}

\begin{figure}[h!]
\begin{center}
\includegraphics{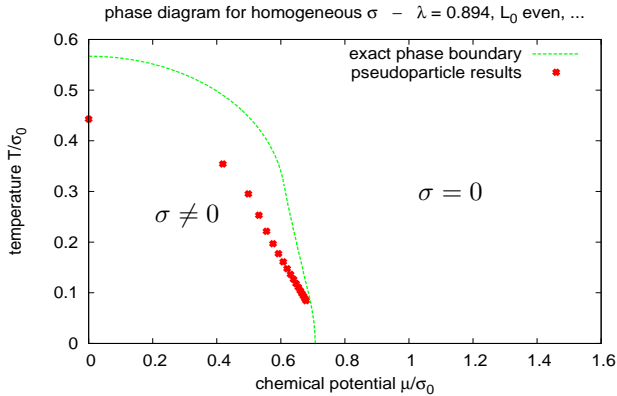}
\caption{\label{FIGURE_005}$Q$-regularization, $\lambda = 0.894$, $L_0$ even, $L_1 = 144$. The pseudoparticle phase diagram for homogeneous chiral condensate together with the exact phase boundary.}
\end{center}
\end{figure}

For odd $L_0$ unphysical pseudoparticle zero modes render the results completely useless: there is no chirally symmetric phase and, therefore, no sensible phase diagram. For even $L_0$ there are no unphysical pseudoparticle zero modes, but still unphysical low lying pseudoparticle modes. The resulting phase diagram is not in quantitative agreement with analytical results \cite{Dashen:1974xz,Wolff:1985av} (cf.\ Fig.~\ref{FIGURE_005}).

\subsection{Spatially inhomogeneous chiral condensate}

In this section we consider a spatially inhomogeneous chiral condensate, i.e.\ $\sigma = \sigma(x_1)$. This gives rise to a new so called crystal phase. In this phase the chiral condensate is given by
\begin{eqnarray}
\label{EQN_015} \sigma(x_1) \ \ = \ \ A \kappa^2 \frac{\textrm{sn}(A x,\kappa) \textrm{cn}(A x,\kappa)}{\textrm{dn}(A x,\kappa)} ,
\end{eqnarray}
where $\textrm{sn}$, $\textrm{cn}$ and $\textrm{dn}$ are Jacobi elliptic functions and $A$ and $\kappa$ are functions of $T$ and $\mu$ \cite{Schnetz:2004vr}.


\subsubsection{The right phase boundary of the crystal phase}

In addition to the fermionic fields we also represent the chiral condensate in terms of B-spline pseudoparticles:
\begin{eqnarray}
\sigma(x_1) \ \ = \ \ \sum_{n=0}^{L_1 / 3-1} \sigma_n B_{n,\textrm{periodic}}^{(2)}(x_1)
\end{eqnarray}
with uniform knot vector $t_j = 3 j$ and $L_1$ a multiple of $3$ (cf.\ Appendix~\ref{SEC_009}). Note that we consider three times as many degrees of freedom in $x_1$-direction for the fermionic fields than for the chiral condensate. The reason is the following: in general, even low lying eigenfunctions of $Q^\dagger Q$ exhibit more oscillations than the chiral condensate $\sigma$; on the other hand, the fermionic pseudoparticles must be able to approximate these low lying eigenfunctions, in order to produce correct results; therefore the representation of the fermionic fields must be finer than the representation of the chiral condensate.

In accordance with Sec.~\ref{SEC_007} we choose $\lambda = 1.153$ and $L_1 = 144$. We determine the right phase boundary of the crystal phase by computing the eigenvalues of the Hessian matrix of the $Q^\dagger Q$-regularized effective action (\ref{EQN_014}) with respect to $\sigma_n$ at $\sigma = 0$, i.e.\
\begin{eqnarray}
H_{n n'} \ \ = \ \ \left.\frac{\partial}{\partial \sigma_n} \frac{\partial}{\partial \sigma_{n'}} S_\textrm{effective}\right|_{\sigma = 0} .
\end{eqnarray}
Negative eigenvalues indicate inhomogeneous perturbations of the chiral condensate $\sigma = 0$, which decrease the effective action. This in turn is a clear sign of a crystal phase. Of course, the argument does not work the other way round: if there are no negative eigenvalues, the chiral condensate is not necessarily vanishing; $\sigma = 0$ may as well be a local minimum of the effective action; however, a recent investigation of the Gross-Neveu model on the lattice strongly suggests that this is not the case \cite{deForcrand:2006ut}.

\begin{figure}[b!]
\begin{center}
\includegraphics{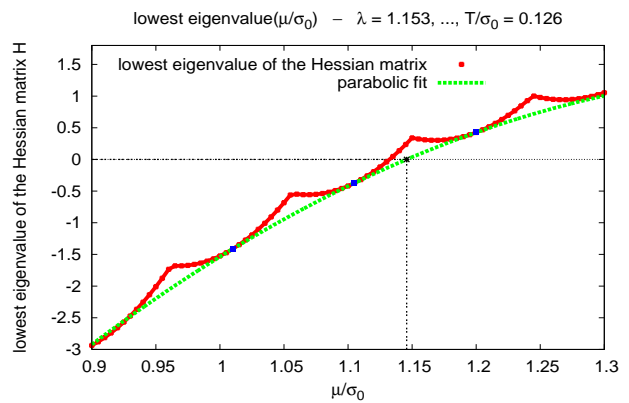}
\caption{\label{FIGURE_006}$\lambda = 1.153$, $L_0 \times L_1 = 36 \times 144$, i.e.\ $T / \sigma_0 = 0.126$. The lowest eigenvalue of the Hessian matrix $H$ as a function of $\mu / \sigma_0$.}
\end{center}
\end{figure}

A minor source of error when computing the right phase boundary is the finite extension of the space dimension. The problem is that the period of the analytically obtained chiral condensate (\ref{EQN_015}) might not ``fit in the periodic $x_1$-direction''. If the period of the chiral condensate is a multiple of the spatial extension $L_1$, everything works fine. However, if it is roughly halfway between two multiples, results differ a little bit (cf.\ Fig.~\ref{FIGURE_006}, where the lowest eigenvalue of $H$ is shown as a function of $\mu / \sigma_0$ [$L_0 = 36$, i.e.\ $T / \sigma_0 = 0.126$]; for a space dimension of infinite extension the graph would be smooth, instead of exhibiting certain periodic oscillations). Note that the same behavior has been observed on the lattice \cite{deForcrand:2006ut}.

\begin{figure}[t!]
\begin{center}
\includegraphics{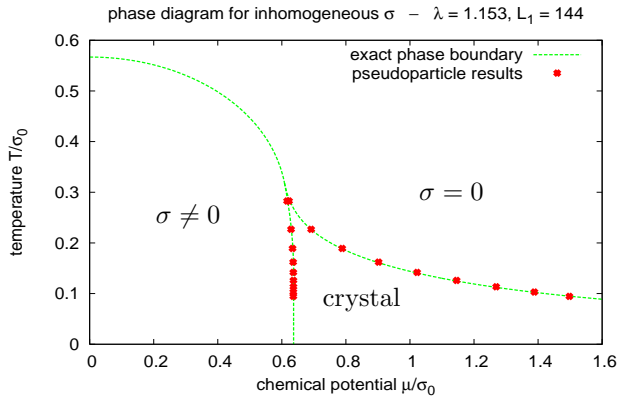}
\caption{\label{FIGURE_007}$Q^\dagger Q$-regularization, $\lambda = 1.153$, $L_1 = 144$. The pseudoparticle phase diagram for spatially inhomogeneous chiral condensate together with the exact phase boundary.}
\end{center}
\end{figure}

To get rid of these finite size effects, we fit a parabola such that it just touches the curve of the lowest eigenvalues from below (cf.\ Fig.~\ref{FIGURE_006}). The root of this parabola is then taken as the corresponding $(\mu / \sigma_0)$-value of the right phase boundary. The resulting right phase boundary is shown in Fig.~\ref{FIGURE_007} together with the analytical result \cite {Schnetz:2004vr}. Up to $\mu / \sigma_0 \approx 1.5$ there is excellent agreement. For larger values of $\mu / \sigma_0$ cutoff effects give rise to certain deviations: the analytically obtained chiral condensate (\ref{EQN_015}) oscillates heavily and, therefore, cannot be represented by $L_1 / 3 = 48$ B-spline pseudoparticles anymore. Of course, applying a larger number of pseudoparticles in $x_1$-direction allows to extract correct results for larger values of $\mu / \sigma_0$.

\subsubsection{The left phase boundary of the crystal phase}

Proceeding in the same way to determine the left phase boundary does not yield correct results. One merely obtains the phase boundary of the phase diagram for homogeneous chiral condensate (cf.\ Fig.~\ref{FIGURE_004}b). The reason is that minima of the effective action for $\sigma = \textrm{constant} \neq 0$ are also minima of the effective action for varying $\sigma$ (however, not necessarily global minima), i.e.\ the Hessian matrix at such values for $\sigma$ is positive definite. To get the correct phase boundary, one has to perform a minimization of the pseudoparticle effective action (\ref{EQN_014}) with respect to $\sigma_n$. Again this is in agreement with what has been observed on the lattice \cite{deForcrand:2006ut}.

The resulting left phase boundary is shown in Fig.~\ref{FIGURE_007}. It is in excellent agreement with the analytical result \cite{Schnetz:2004vr}.


\subsubsection{The chiral condensate}

\begin{figure}[t!]
\begin{center}
\includegraphics{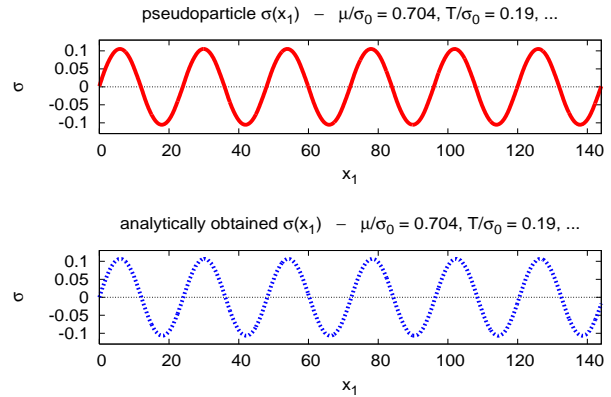}
\caption{\label{FIGURE_008}$\lambda = 1.153$, $\mu / \sigma_0 = 0.704$, $L_0 \times L_1 = 24 \times 144$, i.e.\ $T / \sigma_0 = 0.19$. Pseudoparticle chiral condensate and the corresponding analytically obtained chiral condensate as functions of $x_1$.}
\end{center}
\end{figure}

Another check of the pseudoparticle approach is to compare the chiral condensate obtained by minimizing the $Q^\dagger Q$-regularized effective action (\ref{EQN_014}) with the analytically obtained chiral condensate (\ref{EQN_015}). An example is shown in Fig.~\ref{FIGURE_008} ($\mu / \sigma_0 = 0.704$, $T / \sigma_0 = 0.19$). There is excellent agreement between pseudoparticle and analytical results.

\begin{figure}[h!]
\begin{center}
\includegraphics{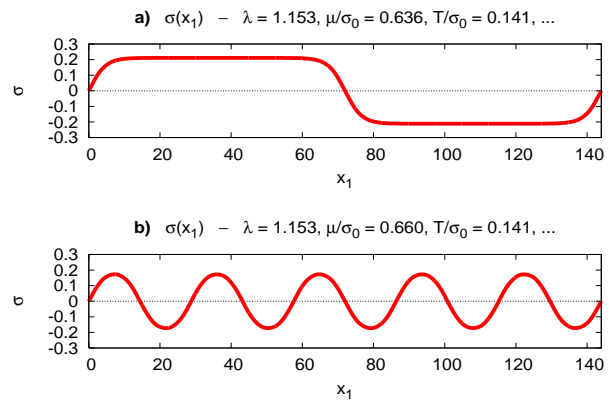}
\caption{\label{FIGURE_009}$\lambda = 1.153$, $L_0 \times L_1 = 32 \times 144$, i.e.\ $T / \sigma_0 = 0.141$. $\sigma$ as a function of $x_1$.
\textbf{a)}~Close to the left phase boundary: $\mu / \sigma_0 = 0.636$.
\textbf{b)}~Inside the crystal phase: $\mu / \sigma_0 = 0.660$.
}
\end{center}
\end{figure}

It is also interesting to compute the chiral condensate for different values of $\mu / \sigma_0$ along a line of constant $T / \sigma_0$. Results are shown in Fig.~\ref{FIGURE_009} as functions of $x_1$ for $T / \sigma_0 = 0.141$. In agreement with \cite{Thies:2003kk,Schnetz:2004vr}, the chiral condensate changes from a sin-like behavior inside the crystal phase (cf.\ Fig.~\ref{FIGURE_009}b) to a kink-antikink structure, when approaching the left phase boundary (cf.\ Fig.~\ref{FIGURE_009}a).

\section{\label{SEC_008}Conclusions and Outlook}

In this paper we have presented a method to incorporate fermions in the pseudoparticle approach.

We have pointed out that a naive pseudoparticle regularization, the $Q$-regularization, is not suited for producing useful numerical results. The problem of this regularization is that applying the Dirac operator to the pseudoparticles yields functions, which are partially outside the pseudoparticle function space. This gives rise to unphysical low lying modes or, in extreme cases, unphysical pseudoparticle zero modes, which spoil the determinant in the effective action.

A slightly different regularization scheme, the $Q^\dagger Q$-regularization, does not suffer from this problem. To test this pseudoparticle regularization, we have computed the phase diagram of the 1+1-dimensional Gross-Neveu model in the large-$N$ limit, both for homogeneous and for spatially inhomogeneous chiral condensate. Our pseudoparticle results are in quantitative agreement with analytical results. We have given a couple of arguments indicating certain relations between the $Q^\dagger Q$-regularization and finite mode regularization, which might explain these excellent results.

The next step is to apply the pseudoparticle approach to QCD. Of course, instead of using a large number of overlapping hat functions as pseudoparticles, as it has been done in this paper, the goal will rather be to use a small number of physically relevant fermionic pseudoparticles, probably pseudoparticles, which have significant overlap to low lying eigenfunctions of the QCD Dirac operator.

Current research includes a study of chiral symmetry breaking in QCD in the pseudoparticle approach. We do this by computing the low lying eigenvalues of the Dirac operator at different temperatures and relate the results to the chiral condensate via the Banks-Casher relation. Of course, the goal is to obtain a model which exhibits chiral symmetry breaking and a confinement deconfinement phase transition at the same time. Such a model could be useful for computing observables, which are difficult to access in lattice calculations, e.g.\ pion masses or decay constants with realistically light quark masses.


\begin{acknowledgments}

It is a pleasure to thank Frieder Lenz for many helpful and inspiring discussions. I also acknowledge useful conversations with Martin Ammon, Felix Karbstein, Jan M.\ Pawlowski, Michael Thies and Konrad Urlichs.

\end{acknowledgments}


\appendix

\section{\label{SEC_009}B-spline basis functions}

B-spline basis functions are piecewise polynomial functions of degree $k$, which are $\mathcal{C}^{k-1}$-continuous:

\newpage

$\quad$
\vspace{-0.5cm}
%
%
\begin{eqnarray}
 & & \hspace{-0.44cm} B_j^{(0)}(x) \ \ = \ \ \left\{\begin{array}{ccc}
1 & \textrm{if} & t_j \leq x < t_{j+1} \\
0 & \textrm{otherwise}
\end{array}\right. \\
\nonumber & & \hspace{-0.44cm} B_j^{(k)}(x) \ \ = \ \ \frac{x-t_j}{t_{j+k}-t_j} B_j^{(k-1)}(x) + \\
 & & \hspace{0.62cm} \frac{t_{j+k+1}-x}{t_{j+k+1}-t_{j+1}} B_{j+1}^{(k-1)}(x)
\end{eqnarray}
with suitably chosen knot vector $\ldots < t_j < t_{j+1} < \ldots$, e.g.\ $t_j = j$ (cf.\ e.g.\ \cite{Fari01,Wolf06}).

\begin{figure}[h!]
\begin{center}
\includegraphics{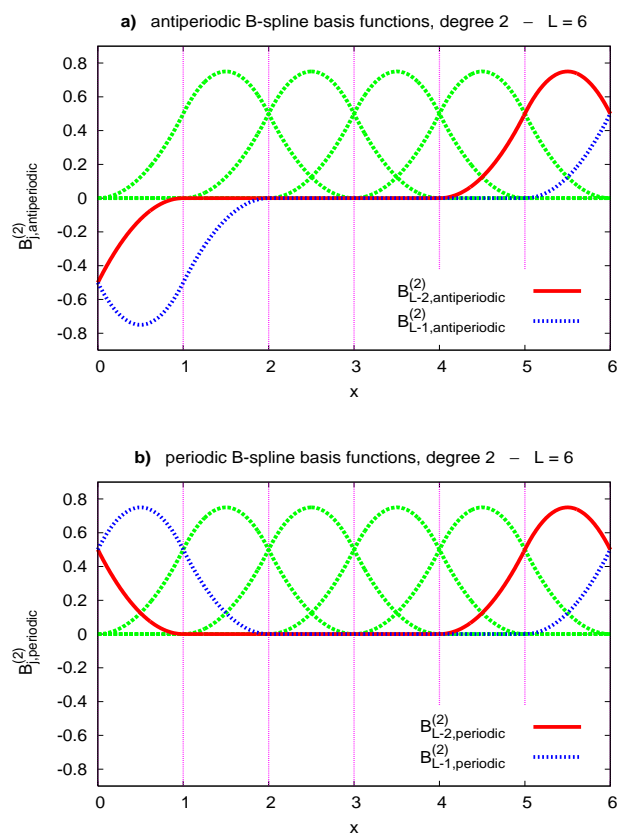}
\caption{\label{FIGURE_010}$L = 6$.
\textbf{a)}~Antiperiodic B-spline basis functions of degree $2$.
\textbf{b)}~Periodic B-spline basis functions of degree $2$.
}
\end{center}
\end{figure}

For $0 \leq x < L$, $L$ integer, antiperiodic and periodic B-spline basis functions with uniform knot vector $t_j = j$ are given by
\begin{eqnarray}
 & & \hspace{-0.44cm} B_{j,\textrm{antiperiodic}}^{(k)}(x) \ \ = \ \ B_j^{(k)}(x) - B_{j-L}^{(k)}(x) \\
 & & \hspace{-0.44cm} B_{j,\textrm{periodic}}^{(k)}(x) \ \ = \ \ B_j^{(k)}(x) + B_{j-L}^{(k)}(x) ,
\end{eqnarray}
$j = 0,\ldots,L-1$. For $L = 6$ and $k = 2$ they are shown in Fig.~\ref{FIGURE_010}.

\begin{thebibliography}{99}

\bibitem{Lenz:2003jp}
  F.~Lenz, J.~W.~Negele and M.~Thies,
  Phys.\ Rev.\ D {\bf 69}, 074009 (2004)
  [arXiv:hep-th/0306105].

\bibitem{Negele:2004hs}
  J.~W.~Negele, F.~Lenz and M.~Thies,
  Nucl.\ Phys.\ Proc.\ Suppl.\ {\bf 140}, 629 (2005)
  [arXiv:hep-lat/0409083].

\bibitem{Wagner:2005vs}
  M.~Wagner and F.~Lenz,
  PoS {\bf LAT2005}, 315 (2006)
  [arXiv:hep-lat/0510083].

\bibitem{Wagner:2006qn}
  M.~Wagner,
  Phys.\ Rev.\ D {\bf 75}, 016004 (2007) \\ $\quad$
  [arXiv:hep-ph/0608090].

\bibitem{Wagner:2006du}
  M.~Wagner,
  AIP Conf.\ Proc.\ {\bf 892}, 231 (2007) \\ $\quad$
  [arXiv:hep-ph/0610291].

\bibitem{Gerhold:2006sk}
  P.~Gerhold, E.~M.~Ilgenfritz and M.~M\"uller-Preussker,
  Nucl.\ Phys.\  B {\bf 760}, 1 (2007)
  [arXiv:hep-ph/0607315].

\bibitem{Gerhold:2006kw}
  P.~Gerhold, E.~M.~Ilgenfritz, M.~M\"uller-Preussker, \\ B.~V.~Martemyanov and A.~I.~Veselov,
  AIP Conf.\ Proc.\ {\bf 892}, 213 (2007)
  [arXiv:hep-ph/0611161].

\bibitem{Andrianov:1982sn}
  A.~A.~Andrianov, L.~Bonora and R.~Gamboa-Saravi,
  Phys.\ Rev.\ D {\bf 26}, 2821 (1982).

\bibitem{Andrianov:1983fg}
  A.~A.~Andrianov and L.~Bonora,
  Nucl.\ Phys.\ B {\bf 233}, 232 (1984).

\bibitem{Andrianov:1983qj}
  A.~A.~Andrianov and L.~Bonora,
  Nucl.\ Phys.\ B {\bf 233}, 247 (1984).

\bibitem{Gross:1974jv}
  D.~J.~Gross and A.~Neveu,
  Phys.\ Rev.\ D {\bf 10}, 3235 (1974).

\bibitem{Dashen:1974xz}
  R.~F.~Dashen, S.~K.~Ma and R.~Rajaraman,
  Phys.\ Rev.\ D {\bf 11}, 1499 (1975).

\bibitem{Wolff:1985av}
  U.~Wolff,
  Phys.\ Lett.\ B {\bf 157}, 303 (1985).

\bibitem{Thies:2003kk}
  M.~Thies and K.~Urlichs,
  Phys.\ Rev.\ D {\bf 67}, 125015 (2003)
  [arXiv:hep-th/0302092].

\bibitem{Schnetz:2004vr}
  O.~Schnetz, M.~Thies and K.~Urlichs,
  Annals Phys.\ {\bf 314}, 425 (2004)
  [arXiv:hep-th/0402014].

\bibitem{Kapu89}
  J.~I.~Kapusta,
  Cambridge University Press (1989).

\bibitem{deForcrand:2006ut}
  P.~de Forcrand and U.~Wenger,
  PoS {\bf LAT2006}, (2006)
  [arXiv:hep-lat/0610117].

\bibitem{Fari01}
  G.~Farin,
  Morgan Kaufmann (2001).

\bibitem{Wolf06}
  \texttt{http://mathworld.wolfram.com/B-Spline.html}.

\end{thebibliography}
\end{document}